\documentstyle[12pt,aps]{revtex}
\begin{document}


\title{
Electromagnetic Energy Penetration in the Self-Induced Transparency Regime
of Relativistic Laser-Plasma Interactions
}
\author{
M.~Tushentsov,$^1$ F.~Cattani,$^2$ A.~Kim,$^1$ D.~Anderson$^2$  and
M.~Lisak$^2$
}
\address{
$^1$Institute of Applied Physics, Russian Academy of Sciences, 603600
Nizhny Novgorod, Russia\\
$^2$Department of Electromagnetics, Chalmers University of Technology,
S-412 96 G\"{o}teborg, Sweden
}
\maketitle

\begin{abstract}

Two scenarios for the penetration of relativistically intense laser radiation
into an overdense plasma, accessible by self-induced transparency,
are presented. For supercritical densities less than 1.5 times the
critical one, penetration of laser energy occurs by
soliton-like  structures moving into the plasma. At higher background
densities laser light penetrates over a finite length
only, that increases with the incident intensity. In this regime
plasma-field structures represent alternating electron layers separated
by about half a wavelength by depleted regions.

\end{abstract}

Recent developments of laser technology have opened possibilities
to explore laser-matter interactions in regimes previously not achievable, \cite{ref}.
This has meant a strong impulse to the theoretical investigation of phenomena occurring
in such extreme conditions, when electrons quiver with relativistic velocities
and new regimes may appear. In particular, penetration of ultra-intense laser
radiation into sharp boundary, overdense plasmas is playing a fundamental role in the
development of
the fast ignitor fusion concept as well as of x-ray lasers, \cite{icf,wilks}. In this regime the optical
properties of the plasma are substantially modified by the
relativistic increase of the inertial electron mass and the consequent lowering of the
natural plasma frequency.\\
In the Seventies it was shown that this relativistic effect enables super-intense
electromagnetic radiation to propagate through classically overdense plasmas, the so
called induced transparency effect, \cite{akhiezer,kaw,marburger,lai}.
Recent numerical simulations based on relativistic PIC codes
\cite{wilks,bonnaud,pukhov,sentoku}, multifluid plasma codes \cite{mason}
and Vlasov simulations \cite{ruhl}, as well as recent experiments
\cite{fuchs,tatarakis}, have revealed a number of new features of the
interaction dynamics, such as laser hole boring, enhanced incident energy
absorption, multi-MeV electron beam, as well as ion beam production and generation of strong magnetic field.\\
An exact analytical study of the stationary stage of the penetration of
relativistically strong radiation into a sharp boundary, semi-infinite, overdense
plasma, taking into account both the relativistic and striction nonlinearity,
has recently led to the determination of an effective threshold intensity for
penetration \cite{fede}. It is known that, for incident intensities lower than the
penetration threshold, an overdense plasma totally reflects the radiation with the
formation of a nonlinear skin-layer structure close to the plasma-vacuum boundary,
\cite{marburger}. For higher intensities the radiation was found to propagate in the
form of nonlinear traveling plane waves \cite{akhiezer,lai}, or
solitary waves  \cite{litvak}. Further analysis has shown that other scenarios are
possible for incident intensities exceeding the threshold, depending on the supercritical plasma
parameter, \cite{ak}. Namely, if $n_o>1.5$ ($n_o$ is the supercritical parameter
defined as $n_o=\omega_{p}^{2}/\omega^2$,
where $\omega$ is the carrier frequency of
the laser, $\omega_{p}=(4\pi e^2N_o/m)^{1/2}$ is the plasma frequency of
the unperturbed plasma), a quasi-stationary state can be
realized and, even if still in a regime of full reflection, the laser energy
penetrates into the overdense plasma over a finite length which depends on the
incident intensity. The subsequent plasma-field structure consists of alternating
electron layers, separated by depleted regions with an extension of about half a wavelength
which
acts as a distributed Bragg reflector. How do these structures emerge as a consequence of relativistic
laser- overdense plasma interactions? What kind of scenarios are realized? These are
the questions we will try to answer in this Letter.\\

Our model is based on relativistic fluid equations for the electrons, in order to
avoid plasma kinetic effects which may
shade or complicate the problem (see, for example, \cite{mason,sudan}). Ions are
considered as a fixed neutralizing background due to the very short time scales
involved, and the slowly varying envelope approximation in time is assumed to be valid.
The governing set of self-consistent
equations for the 1D case of interest in the Coulomb gauge reads
\begin{eqnarray} \label{eq1}
\frac{\partial p_{\parallel}}{\partial t}=\frac{\partial \phi}{\partial x}-
\frac{\partial \gamma}{\partial x}, \\
\frac{\partial n}{\partial t}+\frac{\partial}{\partial
x}(n\frac{p_{\parallel}}{\gamma})=0, \\
\frac{\partial^{2} \varphi}{\partial x^{2}}=n_o (n-1), \\
\label{eq4}
2i\frac{\partial a}{\partial t}+\frac{\partial^2 a}{\partial
x^2}+(1-\frac{n_o}{\gamma}n a)=0.
\end{eqnarray}
Variables are normalized as: $\omega t\rightarrow t$,
$\omega x/c\rightarrow x$, the longitudinal momentum of the electrons
$p_{\parallel}/mc\rightarrow p_{\parallel}$, the scalar potential
$e\varphi/mc^{2}\rightarrow \varphi$, electron density $N/N_o=n$,
  $\gamma=(1+p_{\parallel}^{2}+
a^{2})^{1/2}$ is the Lorentz factor, $m$ and $e$ are the electron rest
mass and charge, $c$ is the speed of light in vacuum and we consider
circularly polarized laser radiation with the amplitude of the
vector potential normalized as $e{\mathbf{A}}/mc^2
=(a(x,t)/\sqrt{2})Re[({\mathbf{y}}+i{\mathbf{z}})\exp(i\omega t)]$.\\
Eqs. (\ref{eq1})-(\ref{eq4}) have been numerically integrated for the problem
of normally incident laser radiation from vacuum ($x<0$) onto a
semi-infinite overdense plasma  ($x\ge 0$), the numerical interval consisting of two
parts: a short vacuum region to the left of the plasma boundary and a semi-infinite
plasma region to the right.\\
As for the boundary conditions, at infinity in the plasma
region the field must vanish, electrons are immobile and the electron density
unperturbed, conditions that are valid until this right boundary is reached by field
perturbations. At the vacuum-plasma boundary the radiation boundary condition reads
\begin{equation}
a-i\frac{\partial a}{\partial x} =2a_{i}(t),
\end{equation}
where $a_{i}(t)$ is the incident laser wave, which means that in the vacuum region the
total field is the sum of the incident and reflected wave.
At the initial time electrons are in equilibrium with ions, i.e., $p_{\parallel}=0,
n=1,\varphi=0$.
Two different cases have been considered for the incident laser pulse: a semi-infinite
envelope turning on as $a_{i}(t)=a_{o}(\tanh t +1)$ and a Gaussian envelope. Finally,
the analysis has been performed for overdense plasmas ($n_o >1$) and for a quite wide
range of incident intensities both higher and lower than the penetration threshold.

For maximum incident intensities lower than the threshold, after a transient stage, a
stationary regime with the formation of nonlinear skin-layers is reached, which is in
perfect agreement with previous analytical solutions \cite{marburger,fede}.
Furthermore,  good agreement is found with the calculated threshold for laser
penetration, \cite{fede}, for intensities above which the nonlinear skin-layer regime
is broken and the interaction leads to the penetration of
laser energy into the overdense plasma. Above this threshold, interactions drastically come into play
and the analysis of this dynamical process, object of a second set of
numerical studies, has revealed two qualitatively different scenarios of laser
penetration into overdense plasmas, depending on the supercritical parameter $n_o$.
Ultimately, the qualitative behavior of the system occurs over a wide range of
incident intensities and thus it does not sensitively depend on the specific values.
If $n_o\leq 1.5$ we have
only a dynamical regime where laser radiation slowly penetrates into the
overdense plasma by moving soliton-like structures. In Fig. (1), the temporal evolution of
the semi-infinite tanh-shaped laser radiation interacting with a plasma with $n_o=1.3$
is depicted.
Solitary waves are generated
near the left boundary and then slowly propagate as quasi-stationary plasma-field
structures
with a velocity much lower than the speed of light. The contribution to the nonlinear
dielectric permittivity due to electron density perturbations is weaker than the one
due to the relativistic nonlinearity, therefore we may consider these solutions as the
extension of pure low-relativistic soliton solutions, \cite{litvak}, to a regime of
slightly higher amplitudes. Furthermore, the excitation dynamics of such
structures is similar to that of structures described by
the nonlinear Schr\"{o}dinger equation with a cubic nonlinearity for a
slightly overdense plasma in the low relativistic limit, (see, i.g., \cite{kolokolov} and references therein).
The generation of similar structures can be inferred from the results of PIC
simulations, such as those presented in \cite{sentoku}.
Thus, if this were the case, i.e., if $n_o -1 \ll 1$, solitary
structures excited by incident intensities slightly above the threshold may be
considered as exact solutions.


When the incident pulse has a Gaussian shape, penetration is seen to occur by a finite
number of soliton-like structures. As shown in Fig. 2(a) a Gaussian pulse with
amplitude $a_o=0.74$ and pulse duration $\tau=200$, for the same plasma parameters as
in Fig. 1, generates two propagating solitary structures instead of a continuous
train. The corresponding spectral analysis, see Fig. 2(b), shows that the spectrum of
the transmitted radiation is on average redshifted, while that of the reflected
radiation presents an unshifted and a blueshifted part which can be accounted for in
terms of Doppler shift due to the moving real vacuum-plasma boundary.\\
It should be underlined that
in the limit of strongly relativistic intensities, when localized solutions have the form
of few-cycle pulses as in \cite{litvak}, our model cannot be applied since
the slowly varying envelope approximation will break down, and the question of what happens at
intensities largely exceeding the threshold is still open.

At higher background densities, $n_o>1.5$, the dynamic regime of interaction is
completely different, as shown in Fig. 3, where a {\em tanh}-like pulse with $a_o=1.3$
that is an intensity of $3.6\times 10^{18} W/cm^2$ for a wavelength of $1 \mu m$
interacts with a plasma with $N_o=1.6 N_{cr}$ ($N_{cr}=m\omega^2/4\pi
e^2$ is the critical density).


The earliest stage of the spatial evolution presents the characteristic distribution
of a nonlinear skin-layer, but the ponderomotive force acting at the vacuum-plasma
boundary is pushing electrons into the plasma, thus shifting the real boundary to a
new position. When the field amplitude on the real boundary exceeds the threshold
calculated in \cite{fede}, the interaction leads to the creation of a deep electron density cavity whose
size is about half a wavelength and which acts as a resonator. The whole plasma-field
structure then starts to slowly penetrate into the plasma and the same process is
repeated at the boundary, where now the perturbed plasma has different
parameters.


 What is interesting is that, after a transient stage during which deep
intensity cavities are produced, the plasma settles down into a quasi-stationary plasma-field
distribution, allowing for penetration of the laser energy over a finite length only,
which increases with increasing incident intensities. The electron density
distribution becomes structured as a sequence of electron layers over the ion
background, separated by about half a wavelength wide depleted regions. The peak
electron density increases from layer to layer reaching an absolute maximum in the
closest layer to the vacuum boundary. At the same time the width of the layers becomes
more and more narrow. Such nonlinear plasma structures can act as a distributed Bragg
reflector and they are very close to those described analytically in \cite{ak}.

If the incident laser pulse has a finite duration, the electromagnetic energy
penetrates into the plasma over a fixed finite length  but, after the laser drive
has vanished, the energy localized inside the plasma is reflected back towards the
vacuum space, as in some sort of ''boomerang'' effect. The transient regime is
obviously more complicated as the depleted regions surrounded by electron layers act
like resonators, with the electromagnetic energy being excited by the incident pulse. Fig.
4 shows how these structures excited by a pulse $400 fs$ long ({\bf
$\lambda=1 \mu m$}) bounce back. Clearly, these excited localized plasma-field
structures may live much longer than the duration time of the drive pulse. However, it
should be underlined that, on a longer time scale, the dynamics can be rather unpredictable. For instance, in a run with
a laser drive $200 fs$ long, the interaction between two structures has resulted in one
long-lived cavity, whereas a second run evolved into a moving localized  structure
similar to those presented in Fig. (1). It is obvious that, when dealing
with long-time dynamics, absorption processes acting on the electromagnetic energy in
the cavities should be taken into account.


In conclusion, we have shown that there are two qualitatively different
scenarios of laser energy penetration into overdense plasmas in the regime
of relativistic self-induced transparency, depending on the background
supercritical density. For slightly supercritical densities $N_o <1.5N_{cr}$,
the penetration of laser energy occurs in the form of long-lived
soliton-like structures which are generated at the vacuum-plasma boundary
plasma and then propagate into the plasma with low velocity. At higher plasma
densities $N_o >1.5N_{cr}$, the interaction results in the generation of
plasma-field structures consisting of alternating electron and electron
displacement regions, with the electromagnetic energy penetrating into the
overdense plasma over a finite length only, as  determined by the incident
intensity.

The work of M.T. and A.K. was supported in part by the Russian Foundation
for Basic Research (grants No. 98-02-17015 and No. 98-02-17013). One of the
authors (F.C.) acknowledges support from the European Community under the
contract ERBFMBICT972428.\\

\pagebreak

FIG. 1. Electron density (solid lines) and field amplitude (dashed lines)
distributions at various moments, for $n_o=1.3$ and semi-infinite pulse
with the maximum incident intensity $a_o=0.74$. The ion ensity distribution
is in dotted line. All the quantities are dimensionless.\\

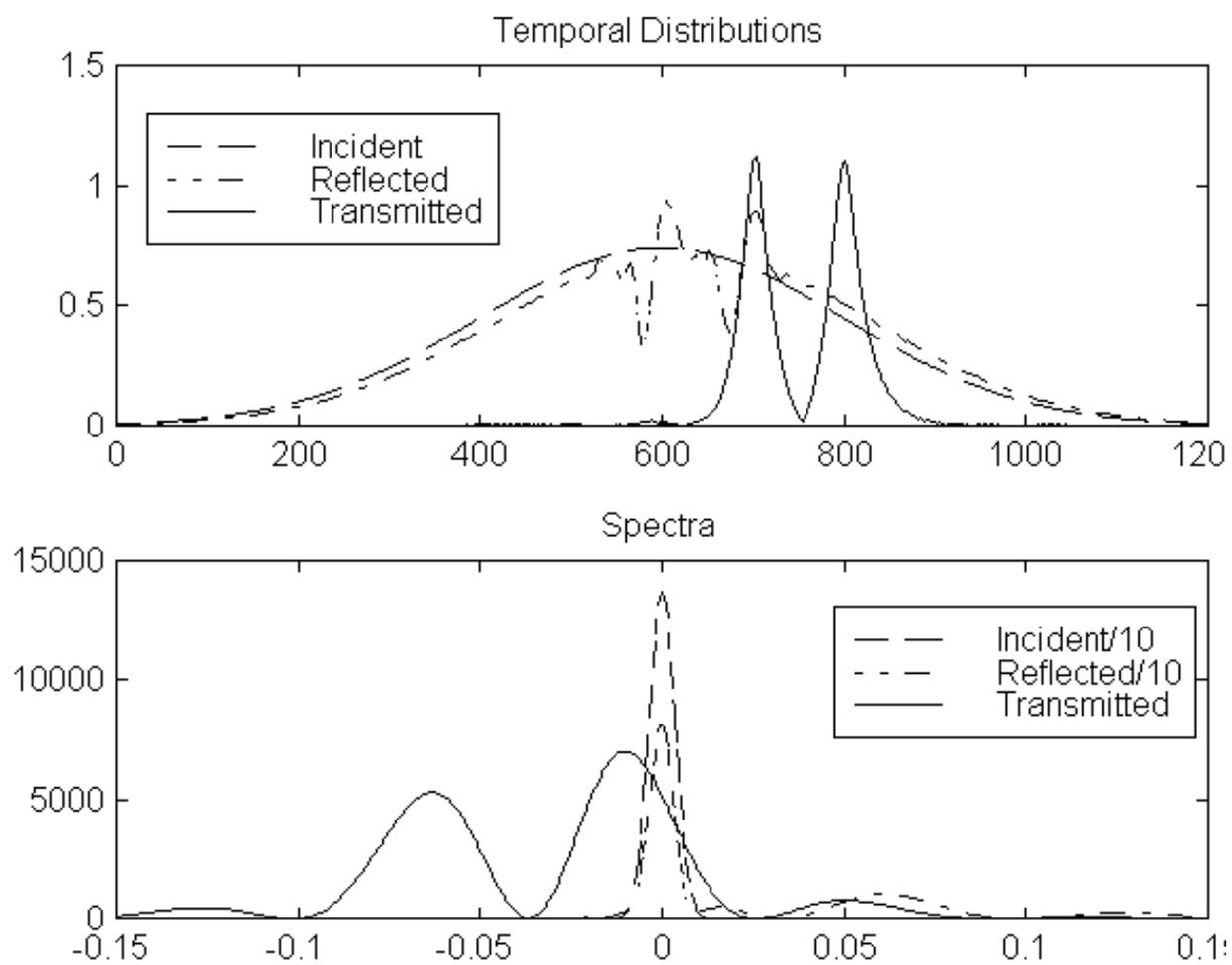
FIG. 2. Temporal distribution (a) of the field structures generated in a plasma with $n_o=1.3$ and $a_{th}=0.62$ by
a Gaussian incident pulse with amplitude $a_o=0.74$ and
width $\tau=200$ and relative spectra (b). All
the quantities are dimensionless.\\

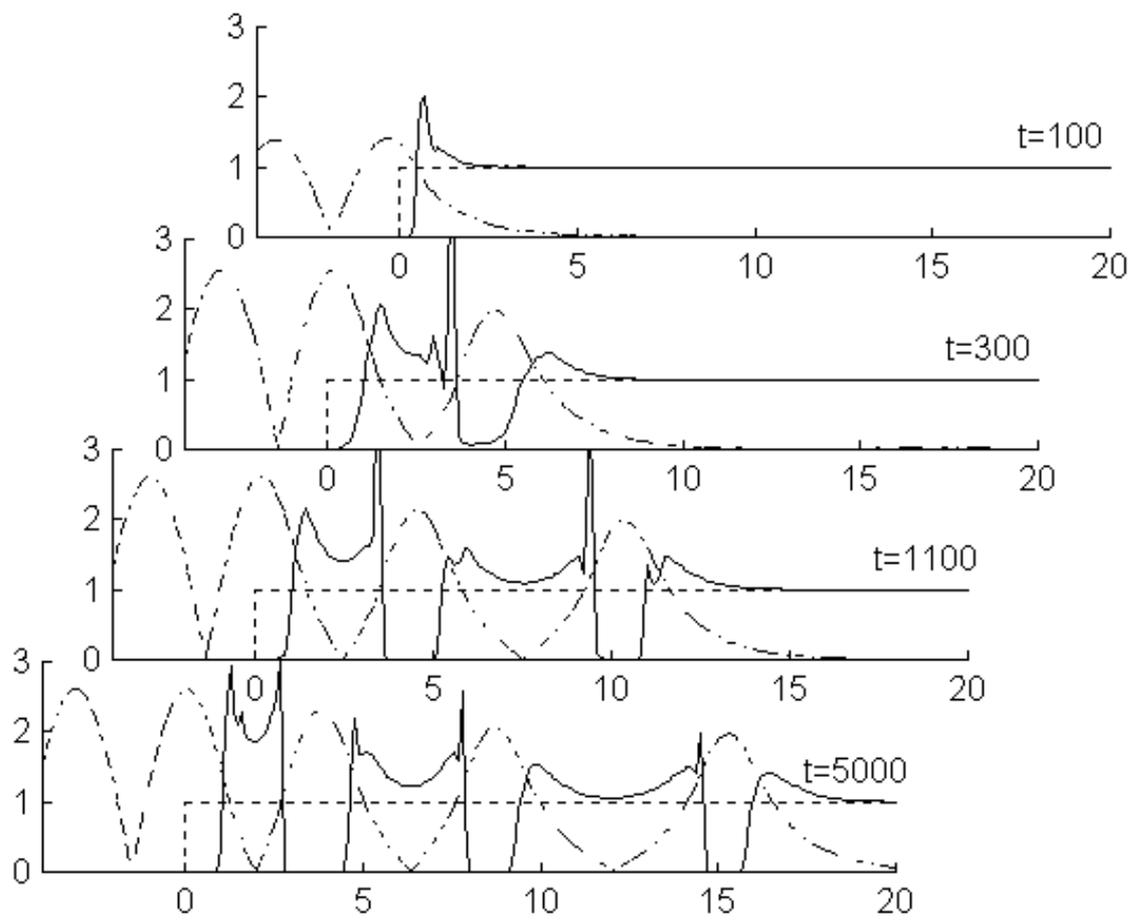
FIG. 3. Snapshots of the time evolution of the electron density (continuous line) and the
solitary structures (dashed line) generated by a semi-infinite
pulse $a_{o}(\tanh t +1$ )with $a_o=1.3$, propagating into a plasma with $n_o=1.6$ and $a_{th}=0.99$.
 All the quantities are dimensionless.\\

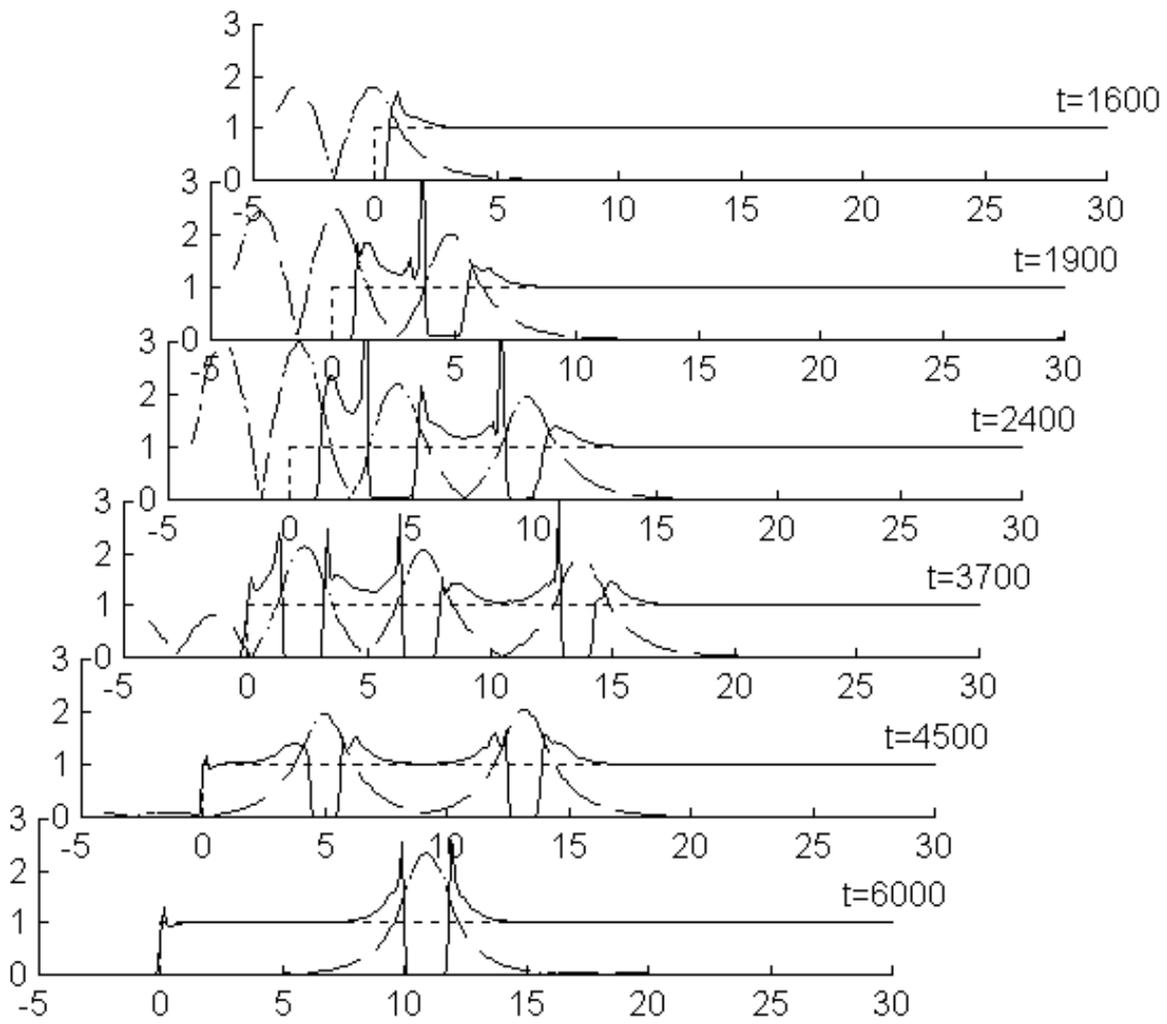
FIG. 4. Temporal distribution of the field structures generated in a plasma with $n_o=1.6$ and $a_{th}=0.99$ by
a Gaussian incident pulse with amplitude $a_o=1.5$ and width $\tau=800$. All the quantities are dimensionless.\\

\pagebreak


\begin{thebibliography}{99}

\bibitem{ref} S.C.~Wilks and W.~Kruer, IEEE Trans. {\bf QE 33}, 154 (1997);
see also in {\em Superstrong Fields in Plasmas},  AIP Conf. Proc. {\bf 426}
(1998).
\bibitem{icf} M.~Tabak {\em et al.}, Phys. Plasmas {\bf 1}, 1626 (1994).
\bibitem{wilks} S.C.~Wilks {\em et al.}, Phys. Rev. Lett. {\bf 69}, 1383
(1992).
\bibitem{akhiezer} A.I.~Akhiezer and R.V.~Polovin, Sov. Phys. JETP {\bf 3},
696 (1956).
\bibitem{kaw} P.~Kaw and J.~Dawson, Phys. Fluids {\bf 13}, 472 (1970);
C.~Max and F.~Perkins, Phys. Rev. Lett. {\bf 27}, 1342 (1971).
\bibitem{marburger} J.H.~Marburger  and R.F.~Tooper, Phys. Rev. Lett. {\bf
35}, 1001 (1975).
\bibitem{lai} C.S.~Lai, Phys. Rev. Lett. {\bf 36}, 966 (1976); F.S.~Felber
and J.H.~Marburger, Phys. Rev. Lett. {\bf 36}, 1176 (1976).
\bibitem{bonnaud} E.~Lefebvre and G.~Bonnaud, Phys. Rev. Lett. {\bf 74},
2002 (1995); H.~Sakagami and K.~Mima, Phys. Rev. E {\bf 54}, 1870 (1996);
S.~Guerin {\em et al.}, Phys. Plasmas {\bf 3}, 2693 (1996); B.F.~Lasinski
{\em et al.}, Phys. Plasmas {\bf 6}, 2041 (1999).
\bibitem{pukhov} A.~Pukhov and J.~Meyer-ter-Vehn, Phys. Rev. Lett. {\bf
79}, 2686 (1997); J.C.~Adam {\em et al.}, Phys. Rev. Lett. {\bf 78}, 4765
(1997).
\bibitem{mason} J.~Mason and M.~Tabak, Phys. Rev. Lett. {\bf 80}, 524 (1998).
\bibitem{ruhl} H.~Ruhl {\em et al.}, Phys. Rev. Lett. {\bf 82}, 2095 (1999).
\bibitem{fuchs} J.~Fuchs {\em et al.}, Phys. Rev. Lett. {\bf 80}, 2326 (1998).
\bibitem{tatarakis} M.~Tatarakis {\em et al.}, Phys. Rev. Lett. {\bf 81},
999 (1998).
\bibitem{fede} F.~Cattani {\em et al.}, Phys. Rev. E {\bf 62}, 1234 (2000).
\bibitem{litvak} V.A.~Kozlov, A.G.~Litvak and E.V.~Suvorov, Sov. Phys.
JETP {\bf 49}, 75, (1979); P.K.~Kaw, A.~Sen and T.~Katsouleas, Phys. Rev.
Lett. {\bf 68}, 3172  (1992).
\bibitem{ak}  A.~Kim  {\em et al.}, JETP Lett.  {\bf 72}, 241 (2000).
\bibitem{sudan} X.L.~Chen and R.N.~Sudan, Phys. Fluids {\bf 5}, 1336 (1993).
\bibitem{kolokolov}  A.V.~Kochetov, Sov. J. Plasma Phys. {\bf 12}, 821 (1986).
\bibitem{sentoku} Y.~Sentoku {\em et al.}, Phys. Rev. Lett. {\bf 83}, 3434
(1999); S.V.~Bulanov {\em et al.}, JETP Lett. {\bf 71}, 407 (2000).

\end{thebibliography}
\end{document}